\documentclass[11pt]{article}
\pdfoutput=1
\usepackage{jheppub}
\usepackage{amsmath}
\usepackage{bm}
\usepackage{slashed}
\usepackage{graphicx}

\newcommand{\tr}{\mathrm{tr}\,}

\newlength{\dummysp}
\newcommand{\beq}{\begin{eqnarray}}
\newcommand{\eeq}{\end{eqnarray}}

\newcommand{\gappeq}{\mathrel{\rlap {\raise.5ex\hbox{$>$}}
{\lower.5ex\hbox{$\sim$}}}}
\newcommand{\lappeq}{\mathrel{\rlap{\raise.5ex\hbox{$<$}}
{\lower.5ex\hbox{$\sim$}}}}

\newcommand{\ben}{\begin{enumerate}}
\newcommand{\een}{\end{enumerate}}

\newcommand{\bit}{\begin{itemize}}
\newcommand{\eit}{\end{itemize}}

\def\[{\left [}
\def\]{\right ]}
\def\({\left (}
\def\){\right )}

\def\Z{{\mathbb Z}}

\title{On the baryon-color-flavor (BCF) anomaly in vector-like theories}
   
 \author[a]{Mohamed M. Anber,}\author[b]{Erich Poppitz} 
  
\affiliation[a]{Department of Physics, Lewis $\&$ Clark College, Portland, OR 97219, USA}
\affiliation[b]{Department of Physics,   University of Toronto, 
Toronto, ON M5S 1A7, Canada}
\emailAdd{manber@lclark.edu}\emailAdd{poppitz@physics.utoronto.ca}    

\abstract{
We consider the most general fractional background fluxes in the color, flavor, and baryon number directions,  compatible with the faithful  action of the global symmetry of a given theory. We call the obstruction to gauging symmetries revealed by such backgrounds the baryon-color-flavor (BCF) 't Hooft anomaly.  We apply the BCF anomaly to vector-like theories, with fermions in higher-dimensional representations of arbitrary N-ality, and derive non-trivial constraints on their IR dynamics. In particular, this class of theories enjoys an independent discrete chiral symmetry and one may ask about the fate of this symmetry in the background of BCF fluxes. We show that, under certain conditions, an anomaly between the chiral symmetry and the BCF background rules out massless composite fermions as the sole player in the IR: either  the composites do not form or additional contributions to the matching of the BCF anomaly are required. We can also give a flavor-symmetric mass to the fermions, smaller than or of order the strong scale of the theory, and examine the $\theta$-angle periodicity of the theory in the BCF background. Interestingly, we find that the conditions that rule out the composites are the exact same conditions that lead to an anomaly of the $\theta$ periodicity: the massive theory will  experience a phase transition as we vary $\theta$ from $0$ to $2\pi$.}

\begin{document}

\maketitle

\flushbottom

\section{Introduction}

The recent interest in higher-form symmetries \cite{Gaiotto:2014kfa} and their 't Hooft anomalies \cite{Gaiotto:2017yup} has resulted in many  applications, see \cite{Gaiotto:2017tne,Anber:2018tcj,Cordova:2018acb,Bi:2018xvr,Wan:2018djl,Anber:2018xek,Anber:2019nfu,Cordova:2019jnf,Cordova:2019uob,Choi:2018tuh,Anber:2018jdf,Bolognesi:2019fej,Kan:2019rsz,Cherman:2019hbq} for a non-comprehensive list. This is not surprising, given that  't Hooft anomaly matching  is one of the very few handles that can give us highly nontrivial information about the infrared (IR) spectrum of strongly coupled theories. 
 
Given a global symmetry $G$ of a theory, we say that the theory has a 't Hooft anomaly if it fails to be gauge invariant as we turn on a background gauge field of $G$. The essence of the new anomaly in \cite{Gaiotto:2014kfa,Gaiotto:2017yup} is that in certain cases the fermions in the theory might not transform faithfully under $G$. For example, if the fermions are charged under a subgroup $\Gamma$ of the center   $\Gamma_c$ of the color group $SU(N_c)$, then the symmetry that acts faithfully on the fermions is $SU(N_c)/\Gamma_r$ (such that the products of all elements in $\Gamma_r$  and $\Gamma$ gives $\Gamma_c$). In this case, one can turn on background gauge fields of  $\Gamma_r$, which carry fractional fluxes and can probe the vacuum of the theory on a finer level.  The fundamental Wilson loops---one-dimensional objects---are charged under $\Gamma_r$, and hence, it is a $1$-form symmetry and the corresponding anomaly is a $1$-form anomaly. 
 
 This is, however, not the only option to probe the theory. For example, theories with  fundamental fermions break center symmetry, and hence, one cannot gauge  center  symmetry or a subgroup of it. Yet, if we have $N_f$ flavors, then the faithful group that acts on the fermions is $SU(N_c)\times SU(N_f)/\mathbb Z_n$, where $n=\mbox{gcd}(N_c,N_f)$. Therefore, one may gauge the center-flavor symmetry $\mathbb Z_n$, which can lead to constraints on the IR dynamics \cite{Shimizu:2017asf}; see also \cite{Tanizaki:2017bam,Tanizaki:2018wtg,Kanazawa:2019tnf} for similar constructions. 
 
In this paper, we turn on the most general fractional background fluxes in the color, flavor, and baryon number directions, which are compatible with the faithful symmetry of a given theory. We call the obstruction to gauging the symmetry the baryon-color-flavor (BCF) 't Hooft anomaly.  We apply BCF anomaly to vector-like theories, with fermions in higher-dimensional representations (arbitrary N-ality), and derive non-trivial constraints on their IR dynamics. 

In particular, this class of theories enjoys a genuine discrete chiral symmetry and one may ask about the fate of this symmetry in the background of BCF fluxes. We show that, under certain conditions, an anomaly between the chiral symmetry and the BCF background rules out massless composite fermions as the sole player in the IR: either  the composites do not form or additional contributions to the matching of the BCF anomaly are required. More specifically, if no integers $\ell^{c,f,B}$ can be found to satisfy (\ref{lattice condition}), the main relation in this paper, then the BCF anomaly is at work; see the bulk of the paper for details. 

We can also give a small flavor-symmetric mass to the fermions, smaller than or of order the strong scale of the theory. Then, we can examine the $\theta$-angle periodicity of the theory in the BCF background. Interestingly, we find that the conditions that rule out the composites are the exact same conditions that lead to an anomaly of the $\theta$ periodicity, see relation (\ref{anomalytheta}): the massive theory will  experience a phase transition as we vary $\theta$ between $0$ and $2\pi$.

This paper is organized as follows. In Section \ref{vector like theories} we review the vector-like theories with fermions in higher-dimensional representations and list their global continuous and discrete symmetries.  In Section \ref{The baryon-color-flavor (BCF) anomaly} we explain the rationale behind the BCF anomaly. We keep our discussion general  since the same anomaly can be extended, with a little extra work, to chiral gauge theories. Then, we show that the BCF anomaly  can lead to non-trivial constraints on the IR dynamics in the vectorlike theories. Next, in Section \ref{BCF anomaly in QCD with massive fermions}, we discuss the role of the new anomaly in massive vector theories with $\theta$-angle and make connections to the anomaly constraints in Section \ref{The baryon-color-flavor (BCF) anomaly}. We conclude in Section \ref{Examples in vector-like theories} by giving some examples and discuss directions for future studies.

\section{Vector-like theories with fermions in higher-dimensional representations}
\label{vector like theories}

We consider $SU(N_c)$ Yang-Mills theory endowed with two sets of left-handed Weyl fermions transforming in representations ${\cal R}^c$ and its complex conjugate $\overline{{\cal R}^c}$ of $SU(N_c)$. In addition, they transform in some general representation ${\cal R}^f$ under their flavor symmetries $SU_L(N_f)\times SU_R(N_f)$. We denote the fermions by $\psi$ and $\tilde \psi$ (both left-handed), respectively, and summarize their gauge and global quantum numbers in the table below:
 \begin{equation}\label{charges1}
\begin{tabular}{|c|c|c|c|c|c|}
\hline
& $SU(N_c)$ & $SU_L(N_f)$ & $SU_R(N_f)$ & $U_B(1)$ & $\mathbb Z_{2\,\mbox{dim}({\cal R}^f)T_{{\cal R}^c}}$\\
\hline
$\psi$ & ${\cal R}^c$ & ${\cal R}^f$ & $1$ & $1$ & $1$\\
\hline
$\tilde\psi$ & $\overline{{\cal R}^c}$ & $1$ & $\overline{{\cal R}^{f}}$ & $-1$ & $1$\\
\hline
\end{tabular}~
\end{equation}
The theory possesses the following  classical global  symmetry
\begin{eqnarray}
SU_L(N_f)\times SU_R(N_f)\times U_L(1)\times U_R(1)\,,
\end{eqnarray}
where $SU_L(N_f)\times U_L(1)$ act on $\psi$, while $SU_R(N_f)\times U_R(1)$ act on $\tilde\psi$.  The colored instantons explicitly  break  $U_L(1)\times U_R(1)$  to  $U_B(1)$ baryon and  discrete chiral symmetries:  $U_L(1)\times U_R(1)\rightarrow U_B(1)\times \mathbb Z_{2\,\mbox{dim}({\cal R}^f)T_{{\cal R}^c}}$, where $\mbox{dim}({\cal R})$ is the dimension of a representation and $T_{\cal R}$ is its Dynkin index (the trace operator) defined as $\mbox{Tr}_{\cal R}\left[T^aT^b\right]=\delta^{ab}T_{\cal R}$ and we normalized our root system such that in the fundamental representation $T_{F}=1$. After modding out the redundant discrete symmetries we find
\begin{eqnarray}
G^{\mbox{global}}=\frac{SU_L(N_f)\times SU_R(N_f)\times U_B(1)\times \mathbb Z_{2\,\mbox{dim}({\cal R}^f)T_{{\cal R}^c}}}{\mathbb Z_{N_f}\times \mathbb Z_{2}}\,. 
\end{eqnarray}
In addition, the theory has a $1$-form center symmetry $\mathbb Z^{(1)}_{p}$ that acts on noncontractible Wilson  loops, provided that $p=\mbox{GCD}(N_c,n_c)>1$, where $n^c$ is the N-ality of ${\cal R}^c$.

One needs to check that $\mathbb Z_{2\,\mbox{dim}({\cal R}^f)T_{{\cal R}^c}}$ is a genuine symmetry of the theory; thus,   it cannot be absorbed in the continuous part of $G^{\mbox{global}}$. Let us assume the converse, namely that    $\mathbb Z_{2\,\mbox{dim}({\cal R}^f)T_{{\cal R}^c}}$, a global rotation that acts on both $\psi$ and $\tilde\psi$, is equivalent to a transformation  combining rotations of $U_B(1)$ and the centers of $SU(N_c)$, $SU_L(N_f)$, and $SU_R(N_f)$:
\begin{eqnarray}
e^{i\frac{2\pi}{2\,\mbox{dim}({\cal R}^f) T_{{\cal R}^c}}}=e^{i\alpha}e^{i\frac{2\pi n_c}{N_c}k_L}e^{i\frac{2\pi n_f }{N_f}p_L}\,, \quad e^{i\frac{2\pi}{2\, \mbox{dim}({\cal R}^f)T_{{\cal R}^c}}}=e^{-i\alpha}e^{-i\frac{2\pi n_c}{N_c}k_R}e^{-i\frac{2\pi n_f }{N_f}p_R}\,,
\end{eqnarray}
where\footnote{Our examples of UV theories will have $n_f=1$, but we keep the discussion general in what follows.} $n_f$ is the N-ality of ${\cal R}^f$ and $k_{L,R}=1,2,..,N_c$, $p_{L,R}=1,2,..,N_f$. The above relations give the condition
\begin{eqnarray}
N_cN_f=\left(n_fN_c(p_L-p_R)+n_cN_f(k_L-k_R)\right)\mbox{dim}({\cal R}^f)T_{{\cal R}^c} + q N_f N_c T_{{\cal{R}}^c}\mbox{dim}({\cal R}^f)\,,
\label{consistency condition}
\end{eqnarray}
for integer $q$. If no solution to (\ref{consistency condition}) exists, then $\mathbb Z_{2\,\mbox{dim}({\cal R}^f)T_{{\cal R}^c}}$ is a genuine (i.e. independent) symmetry of the theory.   Whether (\ref{consistency condition}) has nontrivial solutions can be checked on a case-by-case basis.

\section{The baryon-color-flavor (BCF) anomaly}
\label{The baryon-color-flavor (BCF) anomaly}

In this Section, we sketch the rationale behind the baryon-color-flavor (BCF) anomaly and work out the constraints it imposes on the IR spectrum of a theory. Since this anomaly can be applied to chiral theories as well, we keep our discussion general throughout this section and specialize to the vector-like theories at the end.  

Let $G$ be a  semi-simple Lie group and $\psi$ a fermionic matter field transforming in the irreducible representation ${\cal R}$ of $G$. A quantum field theory of $\psi$ on a manifold ${\cal M}$ is described in terms of a collection of covers $U_{i}$ of ${\cal M}$ along with a set of transition functions $g_{ij} \in G$ on the overlap $U_{ij}=U_i\cap U_j$.  Then,  the matter field, labeled by  $\psi_i$  in the $U_i$ cover, transforms as
 \begin{eqnarray}
 \psi_j=\left(g_{ij}^{\cal R}\right)^{-1}\psi_i
 \label{connection}
 \end{eqnarray}
 on the overlap $U_{ij}$, where $g_{ij}^{\cal R}$ is the transition function of the fields that transform in ${\cal R}$ under $G$. In addition, the transition functions need to satisfy the cocycle condition
\begin{eqnarray}
g_{ij}^{\cal R}g_{jk}^{\cal R}g_{ki}^{\cal R}=1
\label{cocycle condition}
\end{eqnarray}
 on the triple overlap $U_{ijk}=U_{i}\cap U_{j}\cap U_{k}$. If we take $G=SU(N)$ and $g_{ij}$ the transition functions in the defining representation, then we  schematically have 
 \begin{eqnarray}
 g_{ij}^{\cal R}\sim\underbrace{g_{ij}g_{ij}...g_{ij}}_{n\,\mbox{times}}\,,
 \end{eqnarray}
where $n$ is the number of boxes in the Young tableau. However, only the  N-ality of ${\cal R}$, defined as the number of boxes mod $N$, will matter in what follows. From here on we use $n$ to denote the N-ality of ${\cal R}$. If $\mbox{gcd}(N,n)=p$, then there is a subgroup $\mathbb Z_p\subset \mathbb Z_N$, where $\mathbb Z_N$ is  the center   of $SU(N)$,  under which the matter field does not transform. The theory  is invariant under $\mathbb Z_p$: its faithful representation is $SU(N)/\mathbb Z_p$ such that the cocycle condition (\ref{cocycle condition}) can be replaced by the condition
\begin{eqnarray}
g_{ij}g_{jk}g_{ki}=e^{i\frac{2\pi}{p}n_{ijk}}\,
\label{mod cocycle condition}
\end{eqnarray}
on the transition functions of the defining representation and $n_{ijk}$ are integers mod $p$.
 
We can easily generalize the above discussion when $G$ is a direct product of semi-simple Lie groups. We take  $G=SU(N_c)\times SU(N_f)\times U_B(1)$, where $SU(N_c)$, $SU(N_f)$, and $U_B(1)$ are respectively the color, flavor, and baryon number groups ($SU(N_f)$  accounts for both $SU_L(N_f)$ and $SU_R(N_f)$ in vector-like theories). The flavor and baryon number are a subset of the global symmetry of the theory $G^{\mbox{global}}$. We  take $\psi$ to transform in the representation ${\cal R}_c$ under $SU(N_c)$ with N-ality $n_c$ and in the representation ${\cal R}_f$ under $SU(N_f)$ with N-ality $n_f$.  Then, the cocycle conditions (\ref{mod cocycle condition}) are generalized to
\begin{eqnarray}
\nonumber
&&g_{ij}^cg_{jk}^cg_{ki}^c=e^{i\frac{2\pi }{N_c}n_{ijk}^{(c)}}\,,\quad g_{ij}^fg_{jk}^fg_{ki}^f=e^{i\frac{2\pi }{N_f}n_{ijk}^{(f)}}\,,\\
&& g_{ij}^{U_B(1)}g_{jk}^{U_B(1)}g_{ki}^{U_B(1)}=e^{-i n_c\frac{2\pi}{N_c}n_{ijk}^{(c)}-i n_f\frac{2\pi}{N_f}n_{ijk}^{(f)}}\,,
\label{U1 cocycle}
\end{eqnarray}
where $g^c$, $g^f$, and $g^{U_B(1)}$ are the transition functions and $n_{ijk}^{(c)}$ ($n_{ijk}^{(f)}$) are integers mod $N_c$ ($N_f$). Therefore, the group that acts on the fields in the Lagrangian (the faithful representation) is not $SU(N_c)\times SU(N_f)\times U_B(1)$, but rather  $SU(N_c)\times SU(N_f)\times U_B(1)/(\mathbb Z_{N_c}\times\mathbb Z_{N_f})$.

The cocycle conditions (\ref{mod cocycle condition}) and (\ref{U1 cocycle}), as well as  the corresponding faithful representations,  are manifestations of the fact that instantons with integer-valued topological charges will not probe the fine structure of the theory. Turning on fractional instantons, however, can lead to nontrivial constraints on the vacuum structure. In particular, if the theory enjoys one or several discrete global symmetries $\subset G^{\mbox{global}}$, then turning on background fractional fluxes in the color, flavor and baryon number directions (also known as 't Hooft fluxes or twists \cite{tHooft:1979rtg}) can destroy the symmetries. This is a manifestation of a 't Hooft anomaly between the $0$-form global symmetries and the background fluxes. We call this anomaly the ``baryon-color-flavor," or BCF anomaly.

In order to examine the new anomaly, we take  ${\cal M}$ to be a $4$-D torus and turn on appropriate 't Hooft fluxes. Without loss of generality, we consider a square flat torus with cycles of length $L$ and  turn on fluxes  in the $x^1 x^2$ and $x^3 x^4$ planes along the Cartan generators $\bm H$ of $SU(N_c)$, $SU(N_f)$, as well as fluxes in $U_B(1)$. 

To this end, we recall that the kinetic term for a left-handed Weyl fermion $\psi$ that transforms in ${\cal R}_c$ of $SU(N_c)$ and ${\cal R}_f$ of  $SU(N_f)$ is given by: 
\begin{eqnarray}
{\cal L}=i\bar \psi\bar\sigma^\mu\left[\partial_\mu-iA_\mu^cT_{{\cal R}_c}^c-iA_\mu^fT_{{\cal R}_f}^f-i A_\mu^B \right]\psi\,,
\end{eqnarray}
where $T_{{\cal R}_c}^c$ and $T^f_{{\cal R}_f}$ are the Lie-algebra generators of the corresponding groups in the relevant representations. 
We  now turn on 't Hooft fluxes in the $x^1 x^2$ plane by  choosing the following set of gauge field backgrounds, $A_1$ and $A_2$, that are compatible with the cocycle conditions (\ref{U1 cocycle}):
\begin{eqnarray}
\nonumber
A^c_{1}T^c&=&\left(\frac{2\pi m_{12}^c}{L^2}\bf H^c\cdot \bm {\nu}_c \right)x^{2 },\, A^c_2 = 0,\\
\nonumber
A^f_{1 }T^f&=&\left(\frac{2\pi m_{12 }^f}{L^2}\bf H^f\cdot \bm{\nu}_f\right) x^{2 }, \,A^f_2 = 0\\
A^{B}_{1 }&=&\left(n_c\frac{2\pi m^c_{12 }}{L^2N_c}+n_f\frac{2\pi m^f_{12 }}{L^2N_f}\right)x^{2 }\,,A^{B}_{2 }=0. 
\label{gauge fields backgrounds}
\end{eqnarray}
The  generators $T^{c,f}$ ($\bf H^{c,f}$ denote the Cartan subalgebra generators) are taken in the defining representation of the corresponding $SU(N_{c,f})$. Likewise, the weights $\bm \nu_{c,f}$ are  the weights of the fundamental representation of $SU(N_{c,f})$. Furthermore, $m^{c,f}_{12}$ are integers defined mod$(N_{c,f})$ (we ignore an additional allowed integer flux of the baryon background field as it has no effect on our further considerations). The $A_2$ spacetime component of the gauge fields is set to zero by a gauge choice on the 't Hooft fluxes. The backgrounds (\ref{gauge fields backgrounds}) are a particular gauge fixed version of $x^1 x^2$-plane 't Hooft fluxes, useful for practical  and pedestrian\footnote{See Appendix \ref{large GF} for an explicit demonstration that the 't Hooft fluxes shown here are compatible with the cocycle conditions for the transition functions of the $\psi$ fields. We stress that the language of 't Hooft fluxes is not the only one that can be used to reveal  the BCF anomaly. However, it is more widely known than the other formalisms \cite{Gaiotto:2017yup,Gaiotto:2017tne,Shimizu:2017asf,Tanizaki:2017bam,Tanizaki:2018wtg,Wan:2018djl,Cordova:2019uob,Bolognesi:2019fej} and we prefer to use it.}  computations of the anomalies, as  discussed below. 
For brevity, we do not show the 't Hooft fluxes in the $x^3 x^4$ plane: they are obtained from (\ref{gauge fields backgrounds}) by replacing all indices $1 \rightarrow 3$ and $2 \rightarrow 4$, including the integers  $m^{c,f}_{12} \rightarrow m^{c,f}_{34}$ specifying the 't Hooft fluxes.

Using the 't Hooft fluxes (\ref{gauge fields backgrounds}) and their $x^3 x^4$-plane cousins, we find that the corresponding topological charges,  $Q=\frac{1}{16\pi^2}\int_{{\cal M}}F_{\mu\nu}\tilde F_{\mu\nu}$ for the $U(1)$ factors, and $Q = \frac{1}{16\pi^2} \int_{{\cal M}} \tr_{F}\; F_{\mu\nu}\tilde F_{\mu\nu}$ for the nonabelian factors,  are\footnote{We use the identity $\bm \nu_a\cdot \bm\nu_b=\delta_{ab}-\frac{1}{N}$ and we use the same weight in the $12$ and $34$ planes.}
\begin{eqnarray}
\nonumber
Q^c&=&m^c_{12}m^c_{34}\left(1-\frac{1}{N_c}\right)\,, \quad Q^f=m^f_{12}m^f_{34}\left(1-\frac{1}{N_f}\right)\,,\\
Q^{B}&=&\left(n_{c}\frac{m_{12}^c}{N_c}+ n_f\frac{m_{12}^f}{N_f}\right)\left(n_{c}\frac{m_{34}^c}{N_c}+n_f\frac{m_{34}^f}{N_f}\right)\,.
\label{thooft TC}
\end{eqnarray}
As promised,  these topological charges are generically fractional. We stress again that they are not allowed in the original $SU(N_c)$ gauge theory, but serve as nontrivial probes of its dynamics.

Next, we check the condition under which the  BCF anomaly will lead to novel constraints on the IR spectrum of the theory upon applying a global transformation on the matter field. In particular, we examine the situation when the global transformation is a continuous or a genuine discrete symmetry $\subset G^{\mbox{global}}$.

We first review (in a pedestrian way) how the $0$-form 't Hooft anomalies show up as nontrivial phases in the partition function upon performing global transformations on the matter field.  Consider the global transformations $U_g(1)$ and $\mathbb Z_{q_g}$ and take $C^{c}$, $C^{f}$, $C^{B}$  to be the coefficients of the $0$-form anomalies $U_g(1)\left[SU(N_c)\right]^2$, $U_g(1)\left[SU(N_f)\right]^2$, and $U_g(1)\left[U_B(1)\right]^2$, respectively. Similarly, we take $D^{c}$, $D^{f}$, $D^{B}$ to be the anomaly coefficients of $\mathbb Z_{q_g}\left[SU(N_c)\right]^2$, $\mathbb Z_{q_g}\left[SU(N_f)\right]^2$, and $\mathbb Z_{q_g}\left[U_B(1)\right]^2$. 
Then, under  $U_g(1)$ (with parameter $\alpha_g$) and $\mathbb Z_{q_g}$  global transformations the partition function transforms as
\begin{eqnarray}
\nonumber
{\cal Z}|_{U_g(1)}\rightarrow {\cal Z}e^{i\alpha_{g}\left[(1)_c \; C^c +(1)_f \; C^f +(1)_B \; C^B\right]}\,,\quad
{\cal Z}|_{\mathbb Z_{q_g}}\rightarrow {\cal Z}e^{i\frac{2\pi}{q_g}\left[(1)_c\; D^c  + (1)_f\; D^f+ (1)_B\;D^B\right]}\,,\\
\label{PF under ordinary anomalies}
\end{eqnarray}
where $(1)_{c,f,B}$ denotes the smallest value of the color, flavor, and baryon topological charges, which are always integers for $SU(N_{c,f})$ and $U_B(1)$ bundles on the $4$-D torus.
Notice that  the anomalies $U_g(1)\left[SU(N_c)\right]^2$ and $\mathbb Z_{q_g}\left[SU(N_c)\right]^2$ have to vanish, i.e., $C^c=0$ and $D^c=0(\rm{mod} \; q_g$),  since we assumed that both $U_g(1)$ and $\mathbb Z_{q_g}$  are  exact symmetries of the $SU(N_c)$ theory (the reason why we nonetheless kept these coefficients will become clear below).

   The $0$-form 't Hooft anomalies (\ref{PF under ordinary anomalies}) calculated above refer to either the UV or the IR of the theory (one only has to use the appropriate anomaly coefficients). These 't Hooft anomalies must  be matched between the UV and the IR. The matching can occur by Goldstone modes  upon the breaking of $G^{\mbox{global}}$, or by domain walls in the case of breaking a discrete symmetry, or by composite fermions, or by a nontrivial CFT.

Next, in the background of the nontrivial   't Hooft fluxes (\ref{gauge fields backgrounds}), we  replace $(1)_{c,f,B}$ by the fractional topological charges given in (\ref{thooft TC}), and the transformations (\ref{PF under ordinary anomalies})  are replaced with
\begin{eqnarray}
{\cal Z}|_{U_g(1)}\rightarrow {\cal Z}e^{i\alpha_{g}\left[Q^f\; C^f+Q^B\; C^B\right]}\,,\quad
{\cal Z}|_{\mathbb Z_{q_g}}\rightarrow {\cal Z}e^{i\frac{2\pi}{q_g}\left[Q^c\;D^c+Q^f\; D^f+Q^B\;D^B\right]}\,.
\label{PF under new anomalies}
\end{eqnarray}
Here we took into account that $C^c = 0$. Note, however, that while $D^c = 0\,({\rm mod}\, q_g)$, the fractional nature of $Q^c$ for the 't Hooft flux does not allow us to drop it.
Eqns.~(\ref{PF under new anomalies}) represent the new BCF mixed 't Hooft anomalies.

\subsection{Matching by composite fermions}

As was argued by 't Hooft a long time ago, the matching of the $0$-form anomalies in the IR can be achieved via composite fermions.  However, this does not automatically guarantee that the same composites will match the BCF anomaly. In order to show that, we assume the converse, i.e., we assume that there is a set of composites that can solely match the BCF anomaly.  Then,  from the $0$-form anomalies that involve $U_g(1)$, we have:
\begin{eqnarray}
C^{f,B}_{UV}=C^{f,B}_{IR}\,.
\end{eqnarray}
Now, it is easy to see that in this case the first relation in (\ref{PF under new anomalies}), representing the mixed continuous $U_g(1)$-BCF anomaly, gives no new information since the matching of the ``ordinary" continuous 't Hooft anomalies  guarantees that the new anomaly will always be satisfied. 

The anomaly involving $\mathbb Z_{q_g}$ is different since it suffices that the associated anomalies in the UV and IR are matched  mod $q_g$:
\begin{eqnarray}
D^{c,f,B}_{UV}-D^{c,f,B}_{IR}=q_g\ell^{c,f,B}\,, 
\label{the difference}
\end{eqnarray}
and $\ell^{c,f,B}$ are integers. Then, upon substituting (\ref{the difference}) in the second relation in (\ref{PF under new anomalies}) and tracking the anomaly from the UV to the IR we find
\begin{eqnarray}
\frac{{\cal Z}^{UV}|_{\mathbb Z_{q_g}}}{{\cal Z}^{IR}|_{\mathbb Z_{q_g}}}= {e^{i {2 \pi \over q_g} (Q^c D^c_{UV}   + Q^f  D^f_{UV} + Q^B D^B_{UV} ))} \over e^{i {2 \pi \over q_g} (Q^c D^c_{IR}  + Q^F D^f_{IR}   + Q^B D^B_{IR}  )}} = e^{i2\pi \left(\ell^cQ^c+\ell^fQ^f+\ell^BQ^B\right)}\,.
\end{eqnarray}
Thus, if the condition
\begin{eqnarray}
\ell^cQ^c+\ell^fQ^f+\ell^BQ^B \in \mathbb Z\,,
\label{necessary condition}
\end{eqnarray}
is violated for all allowed values of $Q^c$, $Q^a$, and $Q^B$ from (\ref{thooft TC}), then the composites that satisfy the ``ordinary" $0$-form 't Hooft anomalies will fail the BCF anomaly matching. In other words, one needs to show that there exist  integers $\ell^{c,f,B}$ such that the condition (\ref{necessary condition}) is satisfied for the proposed set of composites. 

We can now further simplify the conditions for the integers $\ell^{c,f,B}$ to obey (\ref{necessary condition}). To this end, we first turn on a particular  fractional flux: $Q^B=\frac{n_cn_f}{N_cN_f}, Q^c=Q^f=0$, which is obtained by setting $m_{12}^c=m_{34}^f=1$, $m_{12}^a=m_{34}^c=0$ in (\ref{thooft TC}). Then (\ref{necessary condition}) gives 
\begin{eqnarray}
\ell^B\in {\cal Q}\frac{N_c N_f}{n_cn_f}\mathbb Z\,,
\label{condition on ellB}
\end{eqnarray}
where ${\cal Q}$ is the smallest integer such that  ${\cal Q}\frac{N_c N_f}{n_cn_f}$ is an integer.  Substituting (\ref{condition on ellB}) into (\ref{necessary condition}), and considering now general $m_{12}^{c,f}$, $m_{34}^{c,f}$, we find that (\ref{necessary condition}) is met for general fractional topological charges if  and only if
\begin{eqnarray}
N_c\ell^c-\ell^Bn_c^2\in N_c^2\mathbb Z\,,\quad N_f\ell^f-\ell^B n_f^2\in  N_f^2\mathbb Z\,,\quad \ell^B\in {\cal Q}\frac{N_c N_f}{n_cn_f}\mathbb Z\,.
\label{lattice condition}
\end{eqnarray}
If no integers $\ell^{c,f,B}$  that satisfy (\ref{lattice condition}) can be found, then the BCF anomaly cannot be saturated solely by composite fermions. 

For example, taking $n_f=1$ and $\ell_c=1$,\footnote{To motivate this choice, notice that gauge invariant composites   always have $\ell_c=1$ in the vectorlike theories we consider: this is because, from (\ref{the difference}) it follows that, using $D_{IR}^c = 0$, we have $\ell_c = {D_{UV}^c\over q_g} = {2 T({\cal R}^c) {\rm dim}({\cal R}^f) \over 2 T({\cal R}^c) {\rm dim}({\cal R}^f) } = 1$.} we find that no integers $\ell^{f,B}$  exist if one of the following two conditions is met%
\begin{eqnarray}
\nonumber
&(i)&\,\mbox{gcd}(N_c,N_f)>1\\\
&(ii)&\,\mbox{gcd}(N_c,n_c)>1
\label{conditions for no composites}
\end{eqnarray}
for $N_f\geq 2$ (we have not proven (\ref{conditions for no composites}) but base the conditions upon studying a number of numerical examples).
The conditions (\ref{lattice condition}), (\ref{conditions for no composites})  get modified when we turn off the flavor direction; this is particularly the case if we have a single flavor. We consider examples in Section \ref{Examples in vector-like theories}.

The conditions (\ref{lattice condition}) or (\ref{conditions for no composites}) impose rather  strong constraints on the IR realization of the global symmetries. Interestingly, they are independent of the details of the theory like the  dimension of the representation, the Dynkin index, the global discrete symmetry, etc. They only depend on $N_c$, $N_f$, and the $N$-ality of the representation.  If one of the conditions (\ref{lattice condition}) is not met, then either  the composites do not form or additional contributions to the saturation of the anomalies are required. In this case, the anomaly can be saturated in the IR by (i) CFT, (ii) spontaneous breaking of the global symmetry $\Z_{q_g}$, or (iii)  a topological quantum field theory (TQFT). 

As an example, consider a $SU(N_c)$ vector-like theory with $N_f$ flavors of fermions that transform in a representation ${\cal R}_c$ with N-ality $n_c$ such that $\mbox{gcd}(N_c,N_f)> 1$.  We further assume that the fermions transform in the fundamental of $SU(N_f)$, and hence, $n_f=1$, and that the theory possesses a genuine discrete symmetry. Then, according to the condition (i) in  (\ref{conditions for no composites}) a set of composites  cannot saturate the BCF anomaly. One possibility is that the theory is supplemented by a TQFT. Interestingly,  if we assume $\mbox{gcd}(N_c,n_c)=1$, i.e., there is no center symmetry, then this TQFT is not of a center type. A further investigation of the nature of such TQFTs is beyond the scope of the current work.  

A few simple examples of the constraints are worked out in Section \ref{Examples in vector-like theories}. However, before doing that, we pause in the next  section to discuss an interesting twist obtained upon giving mass to the fermions in the vector-like theories we consider.

\section{BCF anomaly in QCD with massive fermions and nonzero $\mathbf{\theta}$ angle}
\label{BCF anomaly in QCD with massive fermions}

The new anomaly can also put constraints on vector-like theories with non-zero $\theta$ angle and nonzero  flavor-symmetric  fermion masses smaller than,  or of order, $\Lambda$, the strong scale of the theory. Theories with nonzero $\theta$ angle can exhibit anomalies of the $2\pi$ periodicity of $\theta$ if nontrivial backgrounds are turned on, such as our 't Hooft fluxes (\ref{thooft TC}). In this Section, we exhibit a connection between the conditions that composite fermions saturate the BCF anomaly, eqn.~(\ref{lattice condition}) above, and the condition of the absence of anomaly in the $\theta$ periodicity in the massive version of the same theory. We shall see that the two conditions are identical.

  In order to see how the BCF anomaly constrains the massive theory,  consider the most general form of  (the topological part of) the Lagrangian, now including counterterms for the background fields. If the counterterms can be chosen such that the $2 \pi$ periodicity of $\theta$ holds for any background (\ref{thooft TC}), the periodicity anomaly is absent  \cite{Cordova:2019jnf,Cordova:2019uob}. In general color-flavor-baryon number backgrounds, the topological part of the Lagrangian is:
\begin{eqnarray} \label{its}
{\cal L}=\theta Q^c+ \Theta_f  Q^f+ \Theta^B Q^B\,,
\end{eqnarray}
where $Q^c,Q^f,Q^B$ are given by (\ref{thooft TC}) and $\Theta^{f,B}$ are in general real numbers, the coefficients of the baryon and flavor background-field counterterms. We require that these numbers shift by $2\pi$ times integers upon   $2\pi$ shifts of the $\theta$ angle, so  that backgrounds with integer topological charges do not violate the $\theta$ periodicity.
The invariance of the partition function under a shift $\theta\rightarrow \theta+2\pi r$ will be manifest, in arbitrary backgrounds (\ref{thooft TC}), if   integer $s$ and $t$ exist such that  the shift of the Lagrangian  obeys:
\begin{eqnarray}
\Delta{\cal L}=2\pi  r Q^c+2\pi s Q^f+2\pi t Q^B \in 2\pi\mathbb Z\, .
\label{anomaly condition}
\end{eqnarray}
 The condition (\ref{anomaly condition}) is identical to (\ref{necessary condition}). Thus, borrowing the steps that led to (\ref{lattice condition}), we conclude that $\Delta{\cal L}  \in 2\pi \mathbb Z$ is met for general fractional charges  (\ref{thooft TC}) if and only if
\begin{eqnarray}\label{anomalytheta}
N_cr-tn_c^2\in N_c^2\mathbb Z\,,\quad N_fs-t n_f^2\in  N_f^2\mathbb Z\,,\quad t\in {\cal Q}\frac{N_c N_f}{n_cn_f}\mathbb Z\,.
\end{eqnarray}
We note that (\ref{anomalytheta}) generalizes the condition obtained in \cite{Cordova:2019uob} to higher $N$-ality representations; see also \cite{Benini:2017dus} discussing the relation to Chern-Simons theory.

It follows, then, that the condition for the absence of a $\theta$-periodicity anomaly---equivalent to   requiring that (\ref{anomalytheta}) has a  solution  with $r=1$ and integer $s$, $t$---is identical to the condition that massless composites match the BCF anomaly discussed earlier, recall (\ref{lattice condition}) with $\ell_c=1$. This relation reflects the fact that after giving mass to  the vectorlike fermions the anomaly structure of their representations is encoded in the dependence of the IR theory on the phase of the mass parameter.

\section{A simple example of constraints due to BCF anomaly}
\label{Examples in vector-like theories}

We now consider a simple example  of how the BCF anomaly (\ref{PF under new anomalies}) can be used to constrain  the IR phases of QCD-like  theories. We consider ${SU(N_c)}$ gauge theories with one flavor ($N_f=1$) of fermions in the two-index (anti)symmetric ($n_c = 2$) representation.
 The global symmetry of the one-flavor theory is a $\Z_{2 T({\cal R})}$  discrete chiral symmetry\footnote{$T({\cal{R}}) = N_c \pm 2$ for the symmetric ($+$) and antisymmetric ($-$) representation.} and baryon number $U_B(1)$. We now enumerate the various cases:
 \begin{enumerate}
\item $N_c = 4 k +2$, $k>0$. This is the only one flavor two-index (anti-) symmetric theory  whose massive version has an anomaly of its $\theta$-angle periodicity; it is easy to check that the  conditions (\ref{anomalytheta}) have no solutions. Correspondingly, in the massless theory, massless fermion composites  of the form   $\psi^{2 k+1}$ and $\tilde\psi^{2k+1}$ (possibly involving also insertions of field strengths), which can be seen to match all $0$-form 't Hooft anomalies, do not satisfy  BCF anomaly matching (\ref{lattice condition}). 

 The theory also has a $\Z_2^{(1)}$ $1$-form center symmetry which has a mixed $\Z_2$-valued\footnote{\label{center}The topological charge $Q^c$ corresponding to gauging the center equals $N_c\over 4$ here (this was also considered recently in \cite{Bolognesi:2019fej}). This anomaly can be inferred from (\ref{PF under new anomalies}) by taking the fractional part of $Q^c = -{N_c \over 4}$ ($m_{12}^c = m_{34}^c = {N_c \over 2}$)  while $Q^f = Q^b = 0$.} anomaly  with the discrete chiral symmetry. Matching of the $\Z_{2 T({\cal R})}\Z_2^{(1)}$ and BCF anomalies would require (partial) chiral symmetry breaking or an IR TQFT. 
 
\item $N_c = 4 k$.  The massive version of this theory has no $\theta$-periodicity anomaly, as the conditions (\ref{anomalytheta}) can be always satisfied. Fermion composites, however, cannot be constructed here as the baryons are bosons. There is a $\Z_{N-2}$-valued BCF anomaly (\ref{PF under new anomalies})  of the discrete chiral symmetry. As for the other even-$N_c$ case, the theory also has  a $\Z_2^{(1)}$ $1$-form center symmetry but the mixed $\Z_{2 T({\cal R})}\Z_2^{(1)}$ anomaly is absent, see footnote \ref{center}. Thus, it appears that the breaking of the discrete chiral symmetry, along with an IR CFT or a TQFT, are the viable options for matching the anomaly in this theory.

\item $N_c = 2 k + 1$. The massive version of this theory has no $\theta$-periodicity anomaly. Massless ``baryons" of the form $\psi^{2 k+1}$ and $\tilde\psi^{2k+1}$ (with, e.g.,~field strength insertions) can be constructed and shown to match all $0$-form 't Hooft anomalies. Matching of the BCF anomaly is also automatic as per our discussion above. There is no center symmetry in the odd-$N_c$ case and no further anomalies to match. 

It is also possible that anomalies, including the BCF anomaly, are matched by  chiral symmetry breaking. 
\end{enumerate}
One can further consider these theories at higher $N_f$, but we shall not do so here. 
There are other interesting questions arising from the above discussions that are also left for future work.  For example, in   the massive theories with $\theta$-periodicity anomaly (such as our $N_c = 4k+2$ example above) there would have to be   a phase transition at some value of $\theta$ between $0$ and $2\pi$ (presumably $\theta=\pi$ in the CP symmetric case) and one expects 
a  nontrivial structure to arise on domain wall worldvolumes between CP-broken vacua. It would be interesting to study the corresponding domain wall physics.  A related example is of massless theories where the discrete chiral symmetry is broken in the vacuum, as can happen in any of the examples above. These theories are also expected to have  anomaly inflow and it would be interesting to study the associated nontrivial physics on their domain walls (see \cite{Gaiotto:2017tne,Choi:2018tuh} for related work). Figuring out the existence and consistency with the BCF anomaly  of the various TQFTs mentioned above, as well as a detailed study of the $(N_c, N_f, m, \theta)$ phase diagram  is also an interesting task for the future. 
 
{\bf {\flushleft{Acknowledgments:}}} MA  gratefully acknowledges the hospitality at the University of Toronto, where this paper was completed.   MA is supported by the NSF grant PHY-1720135. EP is supported by a Discovery Grant from NSERC and acknowledges the Aspen Center for Physics where some of this work was done.

\appendix 

\section{Compatibility of the 't Hooft fluxes with the matter representation}
\label{large GF}

We consider the $x^1x^2$-plane 't Hooft fluxes from (\ref{gauge fields backgrounds}),  given here for convenience
\begin{eqnarray}
\nonumber
A^c_{1}T^c&=&\left(\frac{2\pi m_{12}^c}{L^2}\bf H^c\cdot \bm {\nu}_c \right)x^{2 },\, A^c_2 = 0,\\
\nonumber
A^f_{1 }T^f&=&\left(\frac{2\pi m_{12 }^f}{L^2}\bf H^f\cdot \bm{\nu}_f\right) x^{2 }, \,A^f_2 = 0\\
A^{B}_{1 }&=&\left(n_c\frac{2\pi m^c_{12 }}{L^2N_c}+n_f\frac{2\pi m^f_{12 }}{L^2N_f}\right)x^{2 }\,,A^{B}_{2 }=0,
\label{gauge fields backgrounds1}
\end{eqnarray}
These gauge backgrounds are not periodic functions on the $x^1 x^2$ torus, since their values at $x^2=L$ and $x^2=0$ are different (recall also that the generators above are taken in the defining representation). However, they are related by gauge transformations. These can be found   from (\ref{gauge fields backgrounds1}), which implies
\begin{eqnarray}
\nonumber  
A^c_{1}T^c(x^2=L)&=&\Omega^{c \; \dagger}(x^1) \left[ A^c_{1}T^c(x^2=0) - i \partial_1\right]\Omega^c(x^1),  ~~~\Omega^c(x^1) = e^{i {2 \pi m_{12}^c x^1 \over L} H^c\cdot \bm {\nu}_c} \\\nonumber
A^f_{1 }T^f(x^2=L)&=&\Omega^{f \; \dagger}(x^1) \left[A^f_{1 }T^f(x^2=0) - i \partial_1\right]\Omega^f(x^1),  ~\Omega^f(x^1) = e^{i {2 \pi m_{12}^f x^1 \over L} H^f\cdot \bm {\nu}_f}\\
A^{B}_{1 }(x^2=L)&=&\Omega^{B \; \dagger} (x^1) \left[A^{B}_{1 }(x^2=0) - i \partial_1\right]\Omega^B(x^1),~~~\Omega^B(x^1) = e^{i {2 \pi  x^1 \over L}({ n_c m_{12}^c \over N_c}+{ n_f m_{12}^f \over N_f})  } \nonumber ~.
\label{gauge fields backgrounds2}
 \end{eqnarray} The gauge group elements $\Omega^{c,f,B}(x^1)$ are the only nontrivial transition functions on the torus in the chosen gauge \cite{tHooft:1979rtg,vanBaal:1982ag}.
We see that $\Omega^{c}(L)$$=$$e^{- i {2 \pi m_{12}^c \over  N_c}} \Omega^c(0)$, $ 
 \Omega^{f}(L)$$=$$e^{- i {2 \pi m_{12}^f  \over  N_f}} \Omega^f(0)$,  $\Omega^B(L)$$=$$e^{i {2 \pi      }({ n_c m_{12}^c \over N_c}+{ n_f m_{12}^f \over N_f})  }\Omega^B(0)$, i.e. they are 
 not periodic functions of $x^1$. However, if a fermion field $\psi$ transforms in a representation of $N$-alities $n_c$ and $n_f$ under $SU(N_c)$ and $SU(N_f)$ (recall that we gave $\psi$ unit charge under $U_B(1)$), the values of $\psi$ at $x^1=0$ and $x^1=L$ are related by the product $(\Omega^{c})^{n_c} (\Omega^{f})^{n_f} \Omega^B$ evaluated at $x^1=L$. This product is clearly unity so that the field $\psi$ is single valued. This provides an explicit demonstration that the background (\ref{gauge fields backgrounds}, \ref{gauge fields backgrounds1}) obeys the cocycle condition (\ref{U1 cocycle}) appropriate to $\psi$ (everything is identically repeated in the $x^3 x^4$-plane).

  \bibliography{References.bib}
  
  \bibliographystyle{JHEP}
  \end{document}